\documentclass[letterpaper, 10 pt, conference]{ieeeconf}
\IEEEoverridecommandlockouts
\usepackage[utf8]{inputenc}
\usepackage{amsmath}
\usepackage{siunitx}
\usepackage{amssymb}
\usepackage{graphicx,balance}
\usepackage{pifont}
\usepackage{bm}
\usepackage{dsfont}
\usepackage{subcaption}
\usepackage{tikz}

\usepackage{amsfonts}
\usepackage{url}
\usepackage{balance}

\usepackage[pdftex,plainpages = false, colorlinks=true, linkcolor=black, citecolor = black, urlcolor = blue,pagebackref=false,hypertexnames=false, plainpages=false, pdfpagelabels]{hyperref}
\usepackage[capitalize]{cleveref}
\usepackage[sort,compress]{cite}

\newtheorem{theorem}{Theorem}[section]

\crefname{section}{Section}{Sections}
\crefname{theorem}{Theorem}{Theorems}
\crefname{lemma}{Lemma}{Lemmas}
\crefname{table}{Table}{Tables}
\crefformat{equation}{(#2#1#3)}
\crefname{algocf}{Algorithm}{Algorithms}
\Crefname{algocf}{Algorithm}{Algorithms}
\crefname{ALC@unique}{Line}{Lines}

\newcommand{\x}{\mathbf{x}}

\newcommand{\w}{\mathbf{w}}
\newcommand{\bu}{\mathbf{u}}
\newcommand{\zz}{\mathbf{z}}

\newcommand{\z}{\mathbf{z}}
\newcommand{\y}{\mathbf{y}}

\newcommand{\etax}{\bm{\eta_x}}
\newcommand{\etav}{\bm{\eta_v}}

\newcommand{\etavt}{\bm{\eta}_{\bm{v},t}}
\newcommand{\etadt}{\bm{\eta}_{des,t}}
\newcommand{\ourbar}[1]{\overline{#1}}

\usepackage{accents}

\usepackage{paralist}

\usepackage[addedmarkup=bf]{changes}
\definechangesauthor[name={MFE}, color={blue}]{MFE}
\definechangesauthor[name={JH}, color={red}]{JH}
\definechangesauthor[name={NR}, color={orange}]{NR}
\crefformat{chapter}{\S#2#1#3}
\crefmultiformat{chapter}{\S\S#2#1#3}{and~#2#1#3}{, #2#1#3}{, and~#2#1#3}
\crefformat{section}{\S#2#1#3}
\crefmultiformat{section}{\S\S#2#1#3}{ and~#2#1#3}{, #2#1#3}{, and~#2#1#3}

\usepackage{algorithm,algorithmic}

\title{\LARGE \bf
 Online Data-Driven Safety Certification for\\Systems Subject to Unknown Disturbances
}

\author{Nicholas Rober$^{*,1}$, Karan Mahesh$^{*,2}$, Tyler M. Paine$^{3,4}$,  Max L. Greene$^{2}$, Steven Lee$^{2}$,\\ Sildomar T. Monteiro$^{2}$, Michael R. Benjamin$^{3}$,  Jonathan P. How$^{1}$%
\thanks{$^{*}$ Indicates equal contribution.}
\thanks{$^{1}$Aerospace Controls Laboratory, Massachusetts Institute of Technology, Cambridge, USA. e-mail: {\tt \small \{nrober,jhow\}@mit.edu}. }
\thanks{$^{2}$Aurora Flight Sciences, a Boeing Company, Cambridge, USA. e-mail: {\tt \small \{mahesh.karan, greene.max, lee.steven, monteiro.sildomar\}@aurora.aero}.} 
\thanks{$^{3}$Marine Autonomy Laboratory, Massachusetts Institute of Technology, Cambridge, USA. e-mail: {\tt \small \{tpaine,mikerb\}@mit.edu}. }
\thanks{$^{4}$Woods Hole Oceanographic Institution, Woods Hole, MA 02543, USA}%
\thanks{This work was supported in part by the US Navy NIWC Atlantic Award N6523623C8011.
This research was developed with funding from the Defense Advanced Research Projects Agency (DARPA). The views, opinions and/or findings expressed are those of the author(s) and should not be interpreted as representing the official views or policies of the Department of Defense or the U.S. Government.
Distribution Statement A. Approved for public release: distribution unlimited.}}

\begin{document}

% \corresp{CORRESPONDING AUTHOR: Nicholas Rober (e-mail: \href{mailto:nrober@mit.edu}{nrober@mit.edu})}
% \authornote{This work was supported in part by Ford Motor Company. The NASA University Leadership Initiative (grant \#80NSSC20M0163) also provided funds to assist the authors with their research. This research was also supported by the National Science Foundation Graduate Research Fellowship under Grant No. DGE–1656518. Any opinion, findings, and conclusions or recommendations expressed in this material are those of the authors and do not necessarily reflect the views of any NASA entity or the National Science Foundation. This work was also supported by AFRL and DARPA under contract FA8750-18-C-0099.}

% \markboth{\MakeUppercase{\thetitle}}{\MakeUppercase{Nicholas Rober} {\itshape ET AL}.}

% \maketitle

% \title{An Important Conference Contribution}

% \author{\IEEEauthorblockN{Author One\IEEEauthorrefmark{1},
% Author Two\IEEEauthorrefmark{2}, Author Three\IEEEauthorrefmark{3} and
% Author Four\IEEEauthorrefmark{4}}
% \IEEEauthorblockA{Department of Whatever,
% Whichever University\\
% Wherever\\
% Email: \IEEEauthorrefmark{1}author.one@add.on.net,
% \IEEEauthorrefmark{2}author.two@add.on.net,
% \IEEEauthorrefmark{3}author.three@add.on.net,
% \IEEEauthorrefmark{4}author.four@add.on.net}}
% \maketitle

\maketitle

\begin{abstract}
Deploying autonomous systems in safety critical settings necessitates methods to verify their safety properties.
This is challenging because real-world systems may be subject to disturbances that affect their performance, but are unknown \emph{a priori}.
This work develops a safety-verification strategy wherein data is collected online and incorporated into a reachability analysis approach to check in real-time that the system avoids dangerous regions of the state space.
Specifically, we employ an optimization-based moving horizon estimator (MHE) to characterize the disturbance affecting the system, which is incorporated into an online reachability calculation. 
Reachable sets are calculated using a computational graph analysis tool to predict the possible future states of the system and verify that they satisfy safety constraints.
We include theoretical arguments proving our approach generates reachable sets that bound the future states of the system, as well as numerical results demonstrating how it can be used for safety verification.
Finally, we present results from hardware experiments demonstrating our approach's ability to perform online reachability calculations for an unmanned surface vehicle subject to currents and actuator failures.
\end{abstract}
\vspace{-1mm}

%!TEX root=main.tex

\section{Introduction}

As autonomous systems are more frequently used in safety-critical settings (e.g., autonomous vehicles~\cite{chen2015deepdriving}, medical diagnosis~\cite{bakator2018deep}, and defense systems~\cite{zheng2021air}), there is a growing need to provide statements about the safety of these systems.
This is especially important as autonomy pipelines become more complicated, possibly including learned components that are sensitive to distribution shifts~\cite{pooch2020can} or perturbations from nominal conditions~\cite{kurakin2016adversarial}.
Generating such safety assurances is challenging because applying them to real-world settings often requires dealing with uncertain, and potentially high-dimensional, systems.
Moreover, in many settings, it is impossible to predict all environmental conditions or system disturbances \emph{a priori}.
In such cases, it is necessary to develop methods capable of certifying safety of a given system at runtime.

Safety assurances can generally be obtained using either testing/data collection~\cite{koopman2016challenges,huang2016autonomous}, or formal analysis~\cite{zhang2018efficient,weng2018towards,xu2020automatic,tjeng2017evaluating,katz2019marabou}.
Testing-based approaches rely on collecting large amounts of data through simulation or deployment of a given system and evaluating if/how the system may fail.
Testing-based approaches are incapable of providing guarantees about safety because it is generally impossible to cover all possible failure cases. 
% Additionally, the deployment of untested safety-critical systems raises policy concerns \cite{vellinga2017testing} and is generally resource intensive, requiring a large number of deployed systems to collect data.
Alternatively, formal methods can provide guarantees about a given system, but their practical application can be limited due to computational constraints or assumptions about the system that may not reflect its actual behavior.
In this paper, we incorporate online data collection into a formal reachability analysis framework to verify the safety of a system, thus striking a balance between data-driven and formal methods.

\begin{figure}[t]
    \centering
    \includegraphics[width=1\columnwidth]{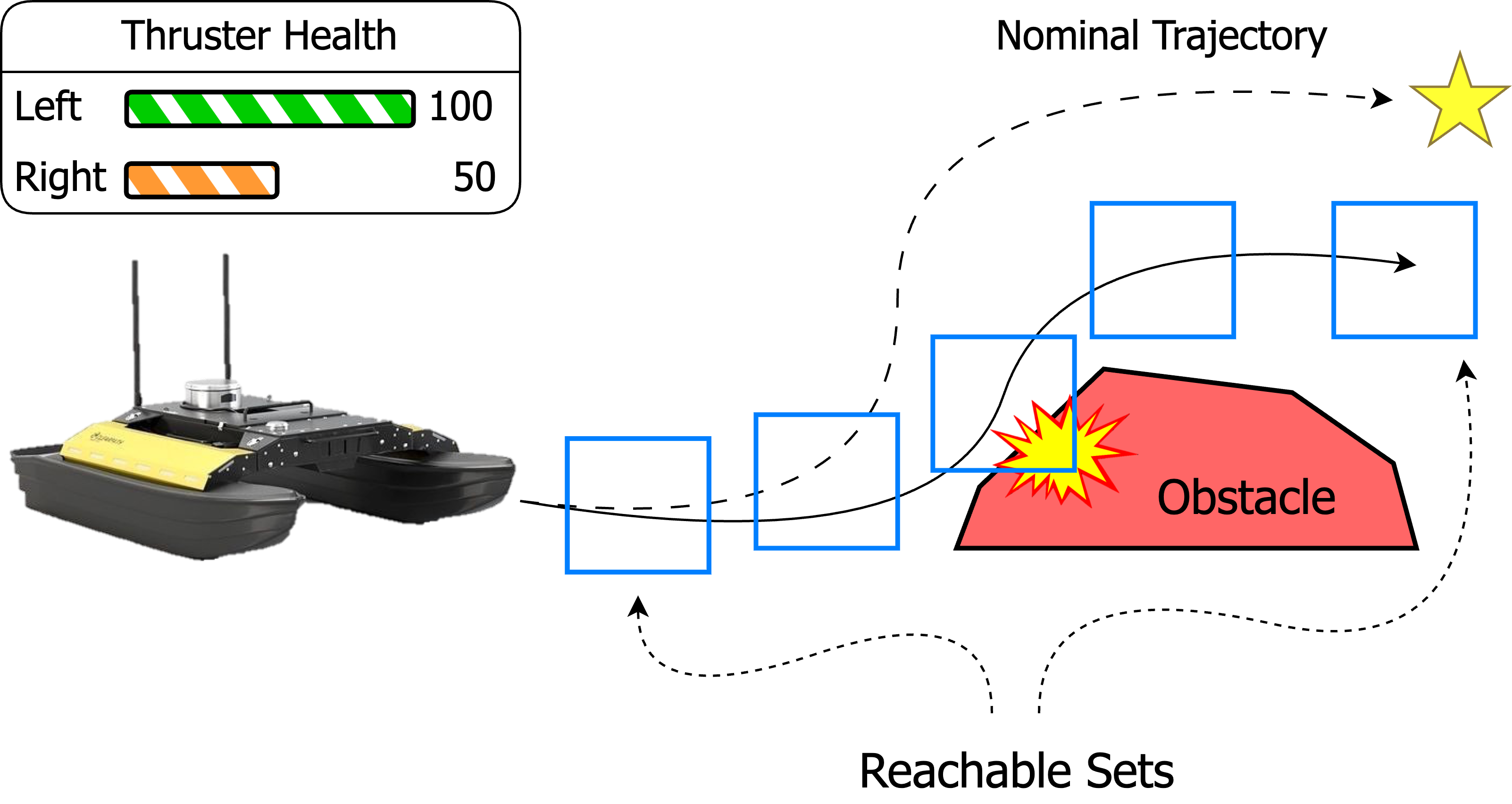}
    \caption{Reachability analysis detects a possible collision for a system experiencing a hardware malfunction}
    \label{fig:intro:boat_reachability_cartoon}
    \vspace{-6mm}
\end{figure}

Reachability analysis, including Hamilton-Jacobi methods~\cite{bansal2017hamilton,evans1998partial}, Lagrangian methods~\cite{gleason2017underapproximation}, and more recent approaches developed for systems with neural networks (NNs)~\cite{vincent2021reachable,dutta2019reachability,huang2019reachnn,ivanov2019verisig,fan2020reachnn,xiang2020reachable,hu2020reach,sidrane2021overt,everett2021reachability}, determines a set of possible future states, i.e., reachable sets (shown in \cref{fig:intro:boat_reachability_cartoon}), of a system given a set of possible initial states.
While all reachability analysis techniques handle uncertainty in the state (reflected in the initial state set), and many consider process/measurement noise~\cite{everett2021reachability,bansal2017hamilton}, they typically assume knowledge of the system's behavior under bounded uncertainty.
Moreover, with only a few exceptions~\cite{lew2021sampling,althoff2014online,herbert2017fastrack}, reachability analysis is often too computationally expensive for online implementation, especially when the system is high-dimensional.
% Other formal approaches, such as safety filtering with control barrier functions (CBFs)~\cite{qin2021learning,dawson2023safe}, have been demonstrated to be learnable and robust to system uncertainty~\cite{dawson2022safe}. However, these results do not consider state uncertainty, which makes generating CBFs more difficult~\cite{dean2021guaranteeing}.

% In this work we leverage a computational graph analysis tool, $\tt{jax\_verify}$~\cite{tip1999practical}, to perform reachability analysis of a system with uncertainty in the state measurements and in the dynamic model.
In this work, we perform online reachability analysis of a system subject to disturbances that are unknown \emph{a priori}. 
Reachable sets are calculated in real time by leveraging $\tt{jax\_verify}$~\cite{jaxverify}, a computational graph analysis tool
% that uses the speed of the $\tt{jax}$ library~\cite{jax2018github} 
capable of efficiently providing output bounds of nonlinear functions.
Disturbances acting on the system are estimated online via a moving horizon estimator (MHE)~\cite{Allgower.Badgwell.ea1999}, and are incorporated into the reachability analysis to predict the behavior of the actual system.
% A window of state data is collected online and fed into a moving horizon estimator (MHE)~\cite{Allgower.Badgwell.ea1999}. The MHE is used to estimate differences between a nominal system model and actual system behavior; these may be due to environmental disturbances, modeling errors, or actuator failures. The difference is estimated by a non-zero-mean noise term, which is incorporated into the reachability analysis to estimate the reachable sets of the actual system.

Our approach is validated using numerical results and deployed on a Clearpath Robotics® Heron unmanned surface vehicle (USV), demonstrating the efficacy of the developed method on a physical system with unknown disturbances.
The main contributions of this paper are as follows:
\begin{itemize}
    % \item The combination of a MHE and reachability for systems with uncertain dynamics, actuator failures, and external disturbances.
    % \item Theoretical arguments supporting the validity of our approach.
    % \item Hardware experiments demonstrating online reachability calculations for an 6DOF autonomous surface vehicle.
    \item Developed a data-driven reachability analysis technique for online safety verification of closed-loop systems subject to disturbances and modeling errors
    \item A proof of the validity of our reachable set over-approximations assuming limits on the rate of change of possible disturbances
    \item Hardware experiments demonstrating real-time reachability calculations at 10 Hz for a 6DOF USV subject to actuator failures and environmental disturbances, e.g., wind and currents
\end{itemize}

% \input{related_work}
%!TEX root=main.tex

\section{Preliminaries}

\subsection{System Dynamics}
Consider the nonlinear system
\begin{equation} \label{eqn:disc_dynamics}
\begin{split}
    & \x_{t+1} = f(\x_{t}, \bu_t) + \w_t \\
    & \y_t = h(\x_t) + \bm{\nu}_t
\end{split}
\end{equation}
where $\x_t \in \mathbb{R}^{n_x}$, $\bu_t \in \mathbb{R}^{n_u}$, $\y_t \in \mathbb{R}^{n_y}$, $\w_t \sim \mathcal{N}(\bm{\mu}_t, \bm{W}_t)$, $\bm{\nu}_t \sim \mathcal{N}(\bm{0}, \bm{R})$ and $t \in \mathbb{N}$.
The dynamics function $f: \mathbb{R}^{n_x} \times \mathbb{R}^{n_u} \rightarrow \mathbb{R}^{n_x}$ and measurement function $h: \mathbb{R}^{n_x} \rightarrow \mathbb{R}^{n_x}$ are assumed to be known, but there is uncertainty in the system due to $\w_t$ because the distribution $\mathcal{N}(\bm{\mu}_t, \bm{W}_t)$ is time-varying and unknown, conditioned on $\bm{W}_t$ being diagonal and 
\begin{equation} \label{eqn:noise_assumption}
    \begin{split}
        & | (\bm{\mu}_{t+1})_i - (\bm{\mu}_t)_i | \leq \Delta\bm{\mu}_i \\
        & | (\bm{\sigma}_{t+1})_i - (\bm{\sigma}_t)_i | \leq \Delta\bm{\sigma}_i,
    \end{split}
\end{equation}
where $(\bm{\sigma}_{t})_i \in \mathbb{R}$ for $i =  1, \ldots , n_x$ is the standard deviation associated with the covariance matrix $\bm{W}_t$, i.e., $(\bm{W}_t)_{ii} = (\bm{\sigma}_t)_i^2$,
%$\bm{\sigma}_t = {\tt sqrt(diag}(\bm{W}_t))$,
and $\Delta\bm{\mu}, \Delta\bm{\sigma} \in \mathbb{R}^{n_x}$ represent the possible variation in $\bm{\mu}_t$ and $\bm{\sigma}_t$, respectively, per time step.
Note that the unknown nature of $\bm{\mu}_t$ and $\bm{W}_t$ (along with the potentially nonlinear dynamics) is a departure from~\cite{everett2021reachability}, where the support of $\w_t$ is assumed known at each time step.
Additionally, while we consider diagonal $\bm{W}_t$ for brevity of our theoretical arguments here, our approach can be extended to the non-diagonal case in future work.
% Consider the sets 
% \begin{equation*} \label{eqn:noise_assumption_wip}
%     \begin{split}
%         S_{t}  &= \{ \min(\sigma(\bm{W}_t)), \max(\sigma(\bm{W}_t)) \} \\
%         S_{t+1} &= \{ \min(\sigma(\bm{W}_{t+1})), \max(\sigma(\bm{W}_{t+1})) \},
%     \end{split}
% \end{equation*}
% where $\sigma \left( \cdot \right)$ is the spectrum, or set of eigenvalues, of the argument.
% We can construct the set $$\Lambda_{t} = \{ \lambda \ \vert\ \lambda = |a-b|,\ \forall a,b \in S_{t}\cup S_{t+1} \}$$ and define $\Delta \lambda := \max(\Lambda)$ as the maximum possible change between.
\let\thefootnote\relax\footnote{ \scriptsize Distribution Statement A. Approved for public release: distribution unlimited. }
Finally, by introducing a general state-feedback control policy $\bu_t = \pi(\x_t)$, the closed-loop dynamics are
\begin{equation} \label{eqn:cl_dynamics}
    \x_{t+1} = f_{cl}(\x_{t}; \pi) + \w_t.
\end{equation}

% but rather there is a known nominal model $\hat{f}: \mathbb{R}^{n_x} \times \mathbb{R}^{n_u} \times \mathbb{R}_{\geq 0} \rightarrow \mathbb{R}^{n_x}$ such that
% \begin{equation} \label{eqn:nominal_assumption}
%     |\hat{f}(\x_{t}, \bu_t, t) + \bm{b}_{t_0} - f(\x_{t}, \bu_t, t)| < \mathbf{w}_{t_0}
% \end{equation}
% holds for all $t \in \mathcal{T} = \{t_0, t_0+1, \ldots, t_0+\tau\}$ where $\bm{b}_{t_0} \in \mathbb{R}^{n_x}$, $\mathbf{w}_{t_0} \in \mathbb{R}^{n_x}$, and $\tau \in \mathbb{R}_{> 0}$.
%The assumption expressed by \cref{eqn:nominal_assumption} specifies that for a given $\tau$, $f$ stays within the tube defined by $\{\x \ \vert\ \x = \hat{f}(\x_{t_0}, \bu_{t_0}, t_0) + \bm{b_{t_0}} + \alpha \mathbf{w}_{t_0}\ \forall \alpha \in [-1, 1]  \}$.

\subsection{Reachability Analysis} \label{sec:prelims:reachability_analysis}
The forward reachable set at time $t+1$ of a system with closed-loop dynamics \cref{eqn:cl_dynamics} is defined recursively as
\begin{equation}
    \mathcal{R}_{t+1}(\mathcal{X}_0) = f_{cl}(\mathcal{R}_t(\mathcal{X}_0); \pi),
\end{equation}
where $\mathcal{R}_0(\mathcal{X}_0) = \mathcal{X}_0$ is the set of possible initial states, and $f_{cl}(\mathcal{R}_t(\mathcal{X}_0); \pi)$ is shorthand for $\{f_{cl}(\x; \pi) \ \vert\ \x \in \mathcal{R}_t(\mathcal{X}_0)\}$.
Going forward, we will omit the $\mathcal{X}_0$ argument unless it is needed for clarity.
% \begin{equation}
%     \mathcal{R}_t(\mathcal{X}_0) = \{\x \ \vert \ f^t (\x_0, \pi(\x_0), 0), \ \x_0 \in \mathcal{X}_0\} 
% \end{equation}
% where $\mathcal{X}_0$ is the set of possible initial conditions, and $f^t$ denotes the $t^\mathrm{th}$ composition of the function $f$, i.e.,
% \begin{align*}
%     & f^t (\x_0, \pi(\x_0), 0) \triangleq f(\x_{t-1}, \pi(\x_{t-1}), t-1) \\ 
%     & \x_{t-1} = f^{t-1}(\x_0, \pi(\x_0), 0) \\
% \end{align*}
% with $f^0(\x_0, \pi(\x_0), 0) = \x_0$.

The exact reachable set $\mathcal{R}_t$ is typically expensive to compute, so we instead compute reachable set over-approximations (RSOAs), i.e., $\ourbar{\mathcal{R}}_t \supseteq \mathcal{R}_t$, as is specified in \cref{sec:approach:data_driven_reach}.
The RSOAs $\{\ourbar{\mathcal{R}}_{t}, \ourbar{\mathcal{R}}_{t+1}, \ldots, \ourbar{\mathcal{R}}_{t+\tau_r}\}$, denoted as $\ourbar{\mathcal{R}}_{t:t+\tau_r}$, can be used to verify safety of the system over a horizon $\tau_r$ by checking if the future states can reach an unsafe region of the state space $\mathcal{C} \subset \mathbb{R}^{n_x}$.
If there is an $ i \in \mathcal{T} = \{t, \ldots, t+\tau_r\}$ such that $\ourbar{\mathcal{R}}_i \bigcap \mathcal{C} \neq \emptyset$, the system may enter the unsafe region and safety is not guaranteed.
Additionally, $\ourbar{\mathcal{R}}_{t:t+\tau_r}$ can be used to check if the system enters a goal region $\mathcal{G}$.
If $\exists i \in \mathcal{T}$ such that $\ourbar{\mathcal{R}}_i \bigcap \mathcal{G} = \ourbar{\mathcal{R}}_i$, the system is guaranteed to reach $\mathcal{G} \subset \mathbb{R}^{n_x}$.
Note that while RSOAs are capable of verifying safety and liveness as described, more conservative RSOAs make the verification conditions harder to satisfy, so tight RSOAs are preferred.

\subsection{Computational Graphs}
\label{sec:prelims:computational_graphs}
A computational graph (CG) $\bm{G}$ is defined as a directed acyclic graph (DAG) with nodes $\bm{V} = \{V_1, V_2, \ldots, V_{n_G}\}$ and edges $\bm{E}$ where each edge is a pair of two nodes $(V_i, V_j)$, indicating that the output of node $V_i$ is an input of node $V_j$.
Each node has an associated computation function $G_i(\cdot)$ consisting of a basic computation such as ReLU or {\tt matmul}, such that $g^{\bm{G}}_i = G_i(u(V_i))$ where $g^{\bm{G}}_i$ is the output of $V_i$ and $u(V_i)$ is the set of inputs to $V_i$, i.e., the outputs from nodes with edges directed toward $V_i$.
The inputs to the graph are denoted as $\z \in \mathbb{R}^{n_i}$.
For brevity going forward, we express $g^{\bm{G}}_i$ in terms of the graph's input, i.e.,  $g^{\bm{G}}_i = g^{\bm{G}}_i(\z)$, thus avoiding the explicit use of $u(V_i)$.
Without loss of generality, we assume $\bm{G}$ has a single output node $V_o$ with dim$(V_o) = \mathbb{R}^{n_o}$, allowing us to express the output of the graph as $g^{\bm{G}}_o(\zz) \in \mathbb{R}^{n_o}$.

\subsection{Computational Graph Relaxation}
\label{sec:prelims:comp_graph_relaxation}
Computational graph relaxation is used to determine relationships between sets of inputs and outputs of a CG.
More specifically, given a set of possible inputs $\mathcal{I}$ to a given CG $\bm{G}$, the goal is to determine a set of possible outputs $\mathcal{O} = \{g^{\bm{G}}_o(\z_0) \ \vert \ \z_0 \in \mathcal{I} \}$.
We construct $\mathcal{I}$ as an $\ell_\infty$-ball, defined as
\begin{align}
    \mathcal{B}_{\infty}(\ourbar{\z}_0, \bm{\epsilon}) &\triangleq \{\z \ \lvert\ \| (\z - \ourbar{\z}_0) \oslash \bm{\epsilon} \|_\infty \leq 1\},
\end{align}
where $\ourbar{\z}_0\in\mathbb{R}^{n_i}$ is the center of the ball, $\bm{\epsilon}\in\mathbb{R}^{n_i}_{\geq 0}$ is a vector whose elements are the radii for the corresponding elements of $\z$, and $\oslash$ denotes element-wise division.

% and specify $\mathcal{I}$ as the set of states within an element-wise $\bm{\epsilon}$-perturbation from $\ourbar{\z}_0$, i.e., 
% \begin{equation*}
% \mathcal{I} = \{\z \ \lvert\ | \z - \ourbar{\z} | \leq \bm{\epsilon}\},
% \end{equation*}
% where $\bm{\epsilon} \in \mathbb{R}^{n_i}$.
% where $\oslash$ represents element-wise division and $\| \cdot \|_\infty$ is the infinity norm.
% \nr{Below is an alternative way to express the equation above. Which do we prefer?}
% \begin{equation}
%     \mathcal{X}_0 = \{\x\ \lvert\ \| (\x - \ourbar{\x}) \oslash \bm{\epsilon} \|_\infty \leq 1\},
% \end{equation}

\begin{theorem}[Linear Relaxation of CGs~\cite{xu2020automatic}] \label{thm:lirpa}
Given a CG $\bm{G}$ and a hyper-rectangular set of possible inputs $\mathcal{I}$, there exist two explicit functions 
\begin{equation*}
    g^{\bm{G}}_{L,o}(\z) = \bm{\Psi}\z + \bm{\alpha},\quad g^{\bm{G}}_{U,o}(\z) =  \bm{\Phi}\z + \bm{\beta}
\end{equation*}
such that the inequality $g^{\bm{G}}_{L,o}(\z) \leq g^{\bm{G}}_o(\z) \leq g^{\bm{G}}_{U,o}(\z)$ holds element-wise for all $\z \in \mathcal{I}$,
with $\bm{\Psi}, \bm{\Phi} \in \mathbb{R}^{n_o \times n_i}$ and $\bm{\alpha}, \bm{\beta} \in \mathbb{R}^{n_o}$.
\end{theorem}
Using \cref{thm:lirpa}, a hyper-rectangular over-approximation of the output set $\mathcal{O}$ is constructed as $$\mathcal{O} \subseteq \ourbar{\mathcal{O}} = \{\bm{o} \ \vert\ g^{\bm{G}}_{L,o}(\z) \leq \bm{o} \leq g^{\bm{G}}_{U,o}(\z),\ \exists \z \in \mathcal{I}\}.$$

\subsection{Moving-Horizon Estimation}
\label{sec:prelims:moving_horizon_est}
MHEs are a class of optimization-based estimators~\cite{Allgower.Badgwell.ea1999}. 
Assuming a system of the form \cref{eqn:disc_dynamics} with a prior on the initial state and its covariance, an MHE uses measurement data from a moving window to estimate the current state of the system and its covariance.
The MHE optimization formulation is constructed as
\begin{equation} \label{eqn:mhe_optim}
    \min_{
        \hat{\x}_{t:t+\tau_e},
        \hat{\w}_{t:t+\tau_e} } J,
\end{equation}
where the cost function $J$ is 
\begin{equation*}
    J = \|\hat{\x}_t - \ourbar{\x}_t\|^2_{\bm{Q}_{t|t-1}^{-1}} + \sum_{k=t}^{t+\tau_e} \| \y_k - h(\hat{\x}_k) \|^2_{\bm{R}^{-1}} + \|\hat{\w}_k\|^2_{\bm{W}_k^{-1}},
\end{equation*}
where $\ourbar{\x}_t$ is the state estimate prior, $\bm{Q}_{t|t-1} \in \mathbb{R}^{n_x \times n_x}$ is the state uncertainty prior, $\y_{t:t+\tau_e}$ represents data measurements  from time $t$ to $t+\tau_e$, and  $$\|\hat{\x}_t - \ourbar{\x}_t\|^2_{\bm{Q}_{t|t-1}^{-1}} \triangleq (\hat{\x}_t - \ourbar{\x}_t)^\top \bm{Q}_{t|t-1}^{-1} (\hat{\x}_t - \ourbar{\x}_t).$$
The result of \cref{eqn:mhe_optim} are values of $\hat{\x}_{t:t+\tau_e}$ and $\hat{\w}_{t:t+\tau_e}$ that optimally estimate the state and disturbance terms over the given window.
Much like model predictive control~\cite{borrelli_bemporad_morari_2017}, the typical idea behind MHE is to execute \cref{eqn:mhe_optim} and collect $\hat \x_t$ at each discrete time step.
Each iteration of the process calculates a new estimate of $\hat{\x}_t$ is generated with \cref{eqn:mhe_optim}, and the priors of the state estimate $\ourbar{\x}_{t+1}$ and covariance $\bm{Q}_{t+1|t}$ are generated for the next time step using the update law 
\begin{equation}
\begin{split}
    & \ourbar{\x}_{t+1} = f_{cl}(\hat{\x}_t; \pi) + \hat{\w}_t \\
    & \bm{Q}_{t|t} = \left(\bm{Q}_{t|t-1}^{-1} + \bm{H}^\top\bm{R}^{-1}\bm{H}\right)^{-1} \\
    & \bm{Q}_{t+1|t} = \bm{A}\bm{Q}_{t|t}\bm{A}^\top + \bm{W}_t,
\end{split}
\end{equation}
where $\bm{A} = \frac{\partial f_{cl}}{\partial \x}\big|_{\x=\hat{\x}_t}$ and  $\bm{H} = \frac{\partial h}{\partial \x}\big|_{\x=\hat{\x}_t}$.

\let\thefootnote\relax\footnote{ \scriptsize Distribution Statement A. Approved for public release: distribution unlimited. }
%!TEX root=main.tex

\section{Reachability for Uncertain Systems} \label{sec:approach:data_driven_reach}
In this section we outline our approach to generating RSOAs for a system of the form \cref{eqn:cl_dynamics}, subject to \emph{a priori} unknown disturbances.
Our approach is designed to be executed online, regularly generating RSOAs over a finite time horizon at a fixed interval.
The proposed data-driven reachability approach is summarized as follows:
\begin{enumerate}
    \item First, before the deployment of the system, we construct $f_{cl}$ as a CG.
    \item At each time step during runtime, we execute an MHE iteration to obtain $\hat{\x}_t$ and estimates of the most recent mean and covariance values of $\w_t$.
    \item Finally, we feed the mean and covariance estimates from the MHE into a CG relaxation to conduct reachability analysis, thus predicting the behavior of the actual system.
\end{enumerate}
% First, before to the deployment of the system, we construct $f_{cl}$ as a CG.
% Then, at each time step during runtime, we execute an MHE iteration to obtain $\hat{\x}_t$ and estimates of the most recent bias and covariance values of $\w_t$.
% Finally, we feed the bias and noise estimates from the MHE into a CG relaxation to conduct reachability analysis, thus predicting the behavior of the actual system.
%
We first address the MHE component.
Notice that the MHE formulation described in \cref{sec:prelims:moving_horizon_est} uses data in the forward direction, but in practice we only have access to data up until time $t$.
Thus, we configure the MHE to use data over the window from time $t-\tau_e$ to time $t$, maintaining a prior on $\ourbar{\x}_{t-\tau_e}$ and a queue of the most recent output measurements $\y_{t-\tau_e:t}$.
With these values, we obtain $\hat{\x}_{t-\tau_e:t}$ and $\hat{\w}_{t-\tau_e:t}$ from \cref{eqn:mhe_optim}.
As described in \cref{sec:prelims:moving_horizon_est}, we collect the state estimate $\hat{\x}_t$ and generate the prior $\ourbar{\x}_{t + 1}$, but we also use the rest of the information obtained from  $\hat{\w}_{t-\tau_e:t}$ to calculate estimates of $\bm{\mu}_t$ and $\bm{W}_t$.
Specifically, we make the approximations $\hat{\bm{\mu}}_t = {\tt mean} (\hat{\w}_{t-\tau_e:t})$ and $\hat{\bm{W}}_t = {\tt cov} (\hat{\w}_{t-\tau_e:t})$.

\begin{figure}[t]
    \centering
    \includegraphics[width=\columnwidth]{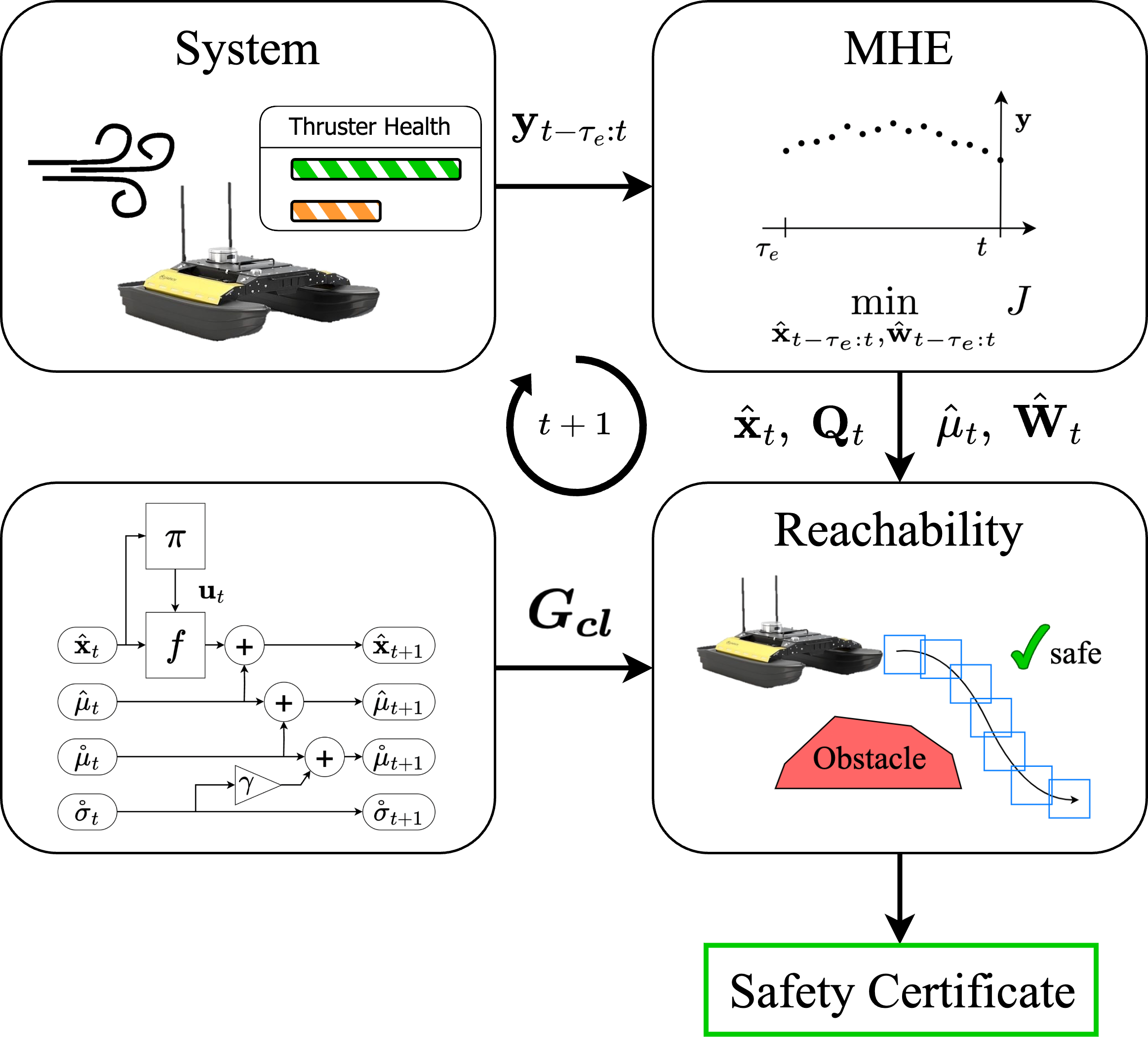}
    \caption{Block diagram depicting our approach. 
    Data is collected from the system and fed into the MHE, which estimates the state and disturbance terms. 
    The outputs of the MHE are used with a CG representation of the closed-loop dynamics to conduct reachability analysis and certify safety.}
    \label{fig:approach:approach_diagram}
    \vspace{-15pt}
\end{figure}

To conduct reachability analysis, we generate RSOAs using the CG analysis tool {\tt jax\_verify}~\cite{jaxverify}.
As shown in \cref{fig:approach:approach_diagram}, by specifying the control policy $\pi$ and the open-loop dynamics \cref{eqn:disc_dynamics} as functions in the {\tt jax\_verify} framework, we can generate a CG representation of \cref{eqn:cl_dynamics} and use \cref{thm:lirpa} to obtain bounds on $\hat{\x}_{t+1}$ from a set of possible $\hat{\x}_t$.
Moreover, to account for the range of possible current and future disturbances, we must also include terms for the disturbance and its rate of change as inputs to the CG.
Thus, we introduce the CG $\bm{G_{cl}}$ with input $\z_t = [\tilde{\x}_t^\top, \tilde{\bm{\mu}}_t^\top, \mathring{\bm{\mu}}_{t}^\top, \mathring{\bm{\sigma}}_{t}^\top]^\top$ and output denoted $g_{o}^{\bm{G_{cl}}}(\z_t) = [\tilde{\x}_{t+1}^\top, \tilde{\bm{\mu}}_{t+1}^\top,  \mathring{\bm{\mu}}_{t+1}^\top, \mathring{\bm{\sigma}}_{t+1}^\top]^\top$ where $\tilde{\x}_t^\top, \tilde{\bm{\mu}}_t^\top,  \mathring{\bm{\mu}}_{t}, \mathring{\bm{\sigma}}_{t} \in \mathbb{R}^{n_x}$.
Note that while the inputs $\tilde{\x}_t$ and $\tilde{\bm{\mu}}_t$ correspond to $\hat{\x}_t$ and $\hat{\w}_{t}$, the inputs $\mathring{\bm{\mu}}_{t}$ and $\mathring{\bm{\sigma}}_{t}$ are internal to the reachability analysis and are used to account for the time variation of $\bm{\mu}_t$ and $\bm{W}_t$ as described by \cref{eqn:noise_assumption}.
$\bm{G_{cl}}$ is then specified by the equations
\begin{align} \label{eqn:graph_update_law:f}
        & \tilde{\x}_{t+1} = f_{cl}(\tilde{\x}_t; \pi) + \tilde{\bm{\mu}}_{t} \\ \label{eqn:graph_update_law:mu}
        & \tilde{\bm{\mu}}_{t+1} = \tilde{\bm{\mu}}_{t} + \mathring{\bm{\mu}}_{t} + \gamma \mathring{\bm{\sigma}}_{t}\\
        %\label{eqn:graph_update_law:sigma}
        %& \tilde{\bm{\sigma}}_{t+1} =  \tilde{\bm{\sigma}}_{t} + \mathring{\bm{\sigma}}_{t}\\ 
        \label{eqn:graph_update_law:delta_mu}
        & \mathring{\bm{\mu}}_{t+1} = \mathring{\bm{\mu}}_{t} \\
        \label{eqn:graph_update_law:delta_sigma}
        & \mathring{\bm{\sigma}}_{t+1} = \mathring{\bm{\sigma}}_{t},
\end{align}
where $\gamma > 0$ is a parameter used to \textit{concretize} an uncertainty bound for a selected confidence interval, e.g., $\gamma=3$ means we assume all samples fall within three standard deviations of the mean.
Concretization is done because {\tt jax\_verify} (and many other analysis tools, such as~\cite{xu2020automatic}) assumes concrete bounds on the possible input states.

% Notice that because we don't have information about the update law for $\bm{\mu}_t$, we set it as constant.
% However, by assuming 
% \begin{equation}
%     | \bm{\mu}_{t+1} - \bm{\mu}_t | \leq \Delta\bm{\mu}
% \end{equation}
% where $\Delta\bm{\mu} \in \mathbb{R}^{n_x}$ represents the possible variation in bias applied to \cref{eqn:cl_dynamics}, we will account for variations in $\bm{\mu}_t$ within the reachability framework as described in the following section.

Having constructed $\bm{G_{cl}}$ and established how we use the MHE, we can now explicitly specify our approach, which is summarized pictorially in \cref{fig:approach:approach_diagram} and via pseudo-code in \cref{alg:safety_cert}.
At each time step, $\y_{t-\tau_e:t}$ is collected from the boat and passed to the MHE, which determines optimal values for $\hat{\x}_{t-\tau_e:t}$ and $\hat{\w}_{t-\tau_e:t}$.
Using the output from the MHE, we determine estimates for the state $\hat{\x}_t$ and its covariance $\bm{Q}_{t|t-1}$, as well as disturbance parameter estimates $\hat{\bm{\mu}}_t$ and $\hat{\bm{W}}_t$.
On \cref{alg:safety_cert:eps_x,alg:safety_cert:eps_mu,alg:safety_cert:eps}, we then obtain concrete uncertainty bounds $\bm{\epsilon}$ by truncating the normal distribution according to concretization parameter $\gamma$ as previously described.
Next we construct the set of possible inputs for $\bm{G_{cl}}$.
On \cref{alg:safety_cert:eps,alg:safety_cert:z,alg:safety_cert:reach_init}, the possible states $\mathcal{B}(\hat{\x}_t, \bm{\epsilon_x})$, noise values $\mathcal{B}(\hat{\bm{\mu}}_t, \bm{\epsilon_\mu})$, and reachability variables $\mathcal{B}({\bm{0}}_6, [\Delta\bm{\mu}^\top,\Delta\bm{\sigma}^\top]^\top)$, are concatenated to get the initial set $\ourbar{\mathcal{R}}_t'$ necessary for reachability analysis.
Next, on \cref{alg:safety_cert:loop,alg:safety_cert:reach_update}
we loop over the horizon $\tau_r$, calculating RSOAs for each time step.
Notice that the RSOA is the set of possible states \textit{and} disturbance terms, $\ourbar{\mathcal{R}}_t' \subset \mathbb{R}^{4n_x}$.
Thus, on \cref{alg:safety_cert:projection}, we project $\ourbar{\mathcal{R}}_t'$ onto $\mathbb{R}^{n_x}$, thereby enabling us to check the safety condition described in \cref{sec:prelims:reachability_analysis} on \cref{alg:safety_cert:safety_check}, which is the desired result.
\begin{algorithm}[t]
 \caption{Online Safety Certification}
 \begin{algorithmic}[1]
 \setcounter{ALC@unique}{0}
 \renewcommand{\algorithmicrequire}{\textbf{Input:}}
 \renewcommand{\algorithmicensure}{\textbf{Output:}}
 \REQUIRE computational graph $\bm{G_{cl}}$, reachability horizon $\tau_r$, vehicle data $\y_{t-\tau_e:t}$, concretization parameter $\gamma$, unsafe region $\mathcal{C}$
 \ENSURE safety certificate $c$ over horizon $\tau_r$
    \STATE $c \leftarrow \mathrm{true}$
    \STATE $\hat{\x}_{t-\tau_e:t}, \hat{\w}_{t-\tau_e:t} \leftarrow \mathrm{MHE}(\y_{t-\tau_e:t})$ \label{alg:safety_cert:mhe}
    \STATE $\bm{\mu}_t \leftarrow {\tt mean}(\hat{\x}_{t-\tau_e:t})$ \label{alg:safety_cert:mean}
    \STATE $\bm{W}_t \leftarrow {\tt cov}(\hat{\w}_{t-\tau_e:t})$  \label{alg:safety_cert:cov}
    \STATE $\bm{\epsilon_x} \leftarrow {\tt concretize}(\mathrm{MHE}.\bm{Q}_t, \gamma)$  \label{alg:safety_cert:eps_x}
    \STATE $\bm{\epsilon_\mu} \leftarrow {\tt concretize}(\bm{W}_t, \gamma)$  \label{alg:safety_cert:eps_mu}
    \STATE $\bm{\epsilon} \leftarrow [\bm{\epsilon_x}^\top, \bm{\epsilon_\mu}^\top, \Delta\bm{\mu}^\top, \gamma \Delta\bm{\sigma}^\top]^\top$ \label{alg:safety_cert:eps}
    \STATE $\z_t \leftarrow [\hat{\x}_t^\top, \hat{\bm{\mu}}_t^\top, \bm{0}_3^\top, \bm{0}_3^\top]^\top$  \label{alg:safety_cert:z}
    \STATE $\ourbar{\mathcal{R}}_t' \leftarrow \mathcal{B}(\z_t, \bm{\epsilon})$  \label{alg:safety_cert:reach_init}
    \FOR{i in $\{t+1,\ \ldots t+\tau_r\}$} \label{alg:safety_cert:loop}
        \STATE $\ourbar{\mathcal{R}}_i' \leftarrow {\tt jax\_verify}(\bm{G_{cl}},\ \ourbar{\mathcal{R}}_{i-1}')$  \label{alg:safety_cert:reach_update}
        \STATE $\ourbar{\mathcal{R}}_i \leftarrow {\tt projection}(\ourbar{\mathcal{R}}_i')$ \label{alg:safety_cert:projection}
        \IF{$\ourbar{\mathcal{R}}_i \bigcap \mathcal{C} \neq \emptyset$}  \label{alg:safety_cert:safety_check}
            \STATE $c \leftarrow \mathrm{false}$ 
        \ENDIF
    \ENDFOR
    
 \RETURN $c$
 \end{algorithmic}\label{alg:safety_cert}
\end{algorithm}
\cref{thm:ours} formally states the result of our approach.
\begin{theorem}[Safety Verification for an Uncertain System] \label{thm:ours}
Consider a system \cref{eqn:cl_dynamics} subject to a disturbance $\w_t \sim \mathcal{N}(\bm{\mu}_t, \bm{W}_t)$ truncated at $\gamma$ standard deviations, where $\bm{\mu}_t$ and $\bm{W}_t$ satisfy the assumptions specified by \cref{eqn:noise_assumption} and where $\hat{\bm{\mu}}_t$ and $\hat{\bm{W}}_t$ are accurate estimates for their respective parameters at the current time step.
The iterative application of \cref{thm:lirpa} with $\bm{G_{cl}}$ defined by \cref{eqn:graph_update_law:f,eqn:graph_update_law:mu,eqn:graph_update_law:delta_mu,eqn:graph_update_law:delta_sigma} and where $\mathcal{I} = \mathcal{B_{\infty}}(\z_t, \bm{\epsilon})$, $\bm{\epsilon} = [\bm{\epsilon_x}^\top, \bm{\epsilon_\mu}^\top, \Delta\bm{\mu}^\top, \gamma\Delta\bm{\sigma}^\top]^\top$ and $\z_t = [\hat{\x}_t^\top, 
\hat{\bm{\mu}}_t^\top, \bm{0}_3^\top, \bm{0}_3^\top]^\top$ provides bounds on all possible $\hat{\x}_{t:t+\tau_r}$, i.e., $\ourbar{\mathcal{R}}_{t:t+\tau_r}$.
\let\thefootnote\relax\footnote{\scriptsize Distribution Statement A. Approved for public release: distribution unlimited.}

\begin{proof}
First, consider the RSOA calculation for time $t+1$.
Since the value of $\w_t$ is bound by a distribution with mean $\hat{\bm{\mu}}_t$ and truncated at $\gamma$ standard deviations with covariance $\hat{\bm{W}}_t$, \cref{eqn:graph_update_law:f} describes all possible values for $\hat{\x}_{t+1}$ when $\tilde{\x}_t \in \mathcal{B}(\hat{\x}_t, \bm{\epsilon_x})$ and $\tilde{\bm{\mu}}_t \in \mathcal{B}(\hat{\bm{\mu}}_t, \bm{\epsilon_\mu})$.
Moreover, since $\hat{\bm{\mu}}_{t+1} \in \mathcal{B}(\hat{\bm{\mu}}_t, \Delta\bm{\mu}+\gamma \Delta\bm{\sigma})$ (via \cref{eqn:noise_assumption}), \cref{eqn:graph_update_law:mu} captures all values of $\hat{\bm{\mu}}_{t+1}$ when $\mathring{\bm{\mu}}_t \in \mathcal{B}(\bm{0}_3, \Delta \bm{\mu})$ and $\mathring{\bm{\sigma}}_t \in \mathcal{B}(\bm{0}_3, \Delta \bm{\sigma})$.
Thus, the application of \cref{thm:lirpa} with $\bm{G_{cl}}$ and $\mathcal{I}$ provides hyper-rectangular bounds on $\hat{\x}_{t+1}$ and $\hat{\bm{\mu}}_{t+1}$, i.e., $\ourbar{\mathcal{R}}'_{t+1} = \{\bm{o} \ \vert\ g^{\bm{G_{cl}}}_{L,o}(\z_t) \leq \bm{o} \leq g^{\bm{G_{cl}}}_{U,o}(\z_t),\ \exists \z_t \in \mathcal{I}\}$.
Let $\ourbar{\mathcal{R}}_{t+1} = \{\x \ \vert\ \x={{proj}_{\mathbb{R}^{n_x}}{\bm{o}}}, \ \forall \bm{o} \in \ourbar{\mathcal{R}}'_{t+1} \}$ (note that because $\ourbar{\mathcal{R}}_{t+1}$ is hyper-rectangular, to project onto $\mathbb{R}^{n_x}$, we simply select the the first ${n_x}$ components of $\ourbar{\mathcal{R}}_{t+1}$).
Since $\ourbar{\mathcal{R}}_{t+1}$ contains the projection of elements of $\ourbar{\mathcal{R}}'_{t+1}$ onto the state-space $\mathbb{R}^{n_x}$, and $\ourbar{\mathcal{R}}'_{t+1}$ provides bounds on $\hat{\x}_{t+1}$ and $\hat{\bm{\mu}}_{t+1}$, $\ourbar{\mathcal{R}}_{t+1}$ is a RSOA for all possible $\hat{\x}_{t+1}$.

To extend this argument to time steps beyond $t+1$, recognize that $\ourbar{\mathcal{R}}'_{t+1}$ gives an \textit{over}-approximation of both $\hat{\x}_{t+1}$ and $\hat{\bm{\mu}}_{t+1}$, allowing us to apply the same argument by considering a new initial state set $\mathcal{I}_1 = \ourbar{\mathcal{R}}'_{t+1}$. 
Thus, by taking over-approximations of over-approximations, we can construct RSOAs over an arbitrary horizon $\tau_r$.
\end{proof}
\end{theorem}

%!TEX root=main.tex

\section{Results}
In this section we demonstrate the performance of our approach with results from both numerical and hardware experiments.
For each set of experiments, we consider a differential-thrust USV whose body-frame velocity vector $\etav = [u, v, r]^\top \in \mathbb{R}^{3}$ characterizes the vehicle's surge $u$, sway $v$, and yaw rate $r$.
We use the widely accepted model of marine vehicle motion ~\cite{fossen2011handbook}:
\begin{equation} \label{eqn:heron_hydrodynamics}
    \bm{M}\dot{\bm{\eta}}_{\bm{v}} + \bm{C}(\bm{M}, \etav)\etav + \bm{D}(\etav)\etav = \bm{\tau}
\end{equation}
where $\bm{M} \in \mathbb{R}^{3 \times 3}$ is the inertia matrix which includes the rigid-body and added mass terms, $\bm{C}(\bm{M}, \etav) \in \mathbb{R}^{3 \times 3}$ is the Coriolis matrix, $\bm{D}(\etav) \in \mathbb{R}^{3 \times 3}$ is the drag matrix, and $\boldsymbol\tau \in \mathbb{R}^{3} $ are the forces and moments acting on the vehicle due to the control inputs.
The position and orientation of the vehicle, $\etax = [x, y, \psi]^\top$, where $x\in \mathbb{R}$ is the $x$-position, $y \in \mathbb{R}$ is the $y$-position, and $\psi\in \mathbb{S}^1$ is the heading angle, are subject to the dynamics
\begin{equation} \label{eqn:heron_kinematics}
    \dot{\bm{\eta}}_{\bm{x}} = \left[
    \begin{aligned}
        u \cos(\psi) - & v \sin(\psi) \\
        u \sin(\psi) + & v \cos(\psi) \\
        r &
    \end{aligned}
    \right].
\end{equation}
For the purpose of control design and application of our approach, we make the following discrete time approximation of the update law for the full USV state $\x = [\etax^\top, \etav^\top]^\top$:
\begin{equation}
    \x_{t+1} = \left[
    \begin{aligned}
         f_{x,d} & (\x_t) \\
        \bm{A}_{cl} \etavt + & \bm{B}_{cl} \etadt
    \end{aligned}
    \right],
\end{equation}
where $f_{x,d}: \mathbb{R}^6 \rightarrow \mathbb{R}^3$ is a discrete approximation of \cref{eqn:heron_kinematics}, $\bm{A}_{cl} \in \mathbb{R}^{3\times3}$ and $\bm{B}_{cl} \in \mathbb{R}^{3\times2}$ characterize the closed-loop dynamics obtained via controller design (discussed further in \cref{sec:results:sim} and \cref{sec:results:hardware}) and system ID techniques~\cite{regan2008rtsm}, and $\etadt = [u_{des,t}, r_{des,t}]^\top \in \mathbb{R}^2$ contains the desired surge $u_{des,t}$ and yaw rate $r_{des,t}$ that the closed-loop system should track.
Note that $\etadt$ is the output of a waypoint-following algorithm, which did not fit into the CG framework required by {\tt jax\_verify}.
Therefore, to predict future values of $\etadt$, we simulate the system forward from the nominal state and collect the desired surges and yaw rates, which are passed to the reachability calculation.

% In \cref{sec:results:sim}

% Rearranging and linearizing \cref{eqn:heron_hydrodynamics} about the operating speed of 1 meter per second, the Heron USV continuous-time MIMO dynamics are
% \begin{equation} \label{eqn:heron_cont_dynamics}
%     \dot\x_{p} = \mathbf{A}_{p}\x_{p} + \mathbf{B}_{p}\bu,
% \end{equation}
% where $\x_{p} \in \mathbb{R}^{3}$  is the concatenated speed (such that speed is the Euclidean norm of surge and sway velocities), sway velocity, and yaw rate state vector, respectively; $\bu \in \mathbb{R}^{2}$ is the concatenated throttle and virtual rudder control inputs; $\mathbf{A}_{p} \in \mathbb{R}^{3 \times 3}$ is the state transition matrix; and $\mathbf{B}_{p} \in \mathbb{R}^{3 \times 2}$ is the control effectiveness matrix.
 
\subsection{Numerical Experiments} \label{sec:results:sim}
First, we use a simulated environment to evaluate the efficacy of our approach while maintaining control over the disturbances that influence the system.
We employ a model reference adaptive controller (MRAC)~\cite{Lavretsky.Wise2012_2}, which provides a control signal to \cref{eqn:heron_hydrodynamics} such that the closed-loop system behaves according to a user-selected $\bm{A}_{cl}$ and $\bm{B}_{cl}$ by design.
The details on the implementation of the closed-loop system and MRAC controller can found in the linked repository.
\subsubsection{Symmetric Thruster Failure}
\label{sec:results:sim:sym}
\begin{figure}[t]
    \centering
    \begin{tikzpicture}
        \usetikzlibrary{decorations.pathreplacing}
        \pgfmathsetmacro{\yshift}{-25px}
        \pgfmathsetmacro{\xshift}{18px}
        \node[anchor=south west,inner sep=0] at (0,0) {\includegraphics[width=\columnwidth, clip, trim={5 10 30 20}]{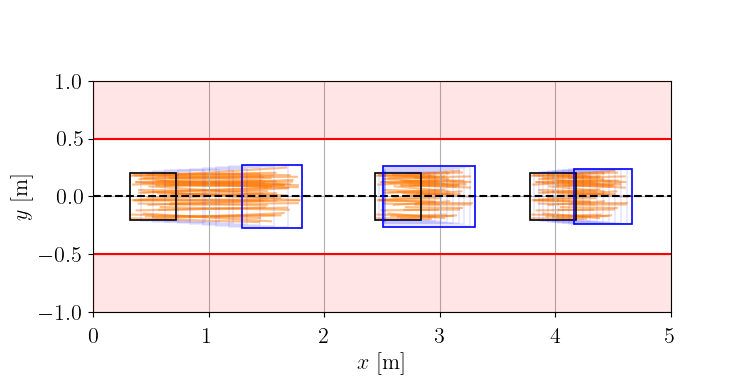}};
        \draw [thick, decorate,decoration={brace,amplitude=10pt}, yshift=\yshift, xshift=\xshift] (0.6,6.2) -- (7.5,6.2) node [black,midway,yshift=16pt] {\small Thruster Health};
        \draw[black,rounded corners,fill=white, yshift=\yshift, xshift=\xshift] (0.6,4.75) rectangle (3.0,6.1);
        \node[text width=1cm, text height=1cm, yshift=\yshift, xshift=\xshift] at (2.28,6.25) {\footnotesize $t=0$};
        \node[text width=1cm, text height=1cm, yshift=\yshift, xshift=\xshift] at (1.22,5.83) {\footnotesize left};
        \node[text width=1cm, text height=1cm, yshift=\yshift, xshift=\xshift] at (1.22,5.4) {\footnotesize right};
        \draw[black,rounded corners,densely dotted, yshift=\yshift, xshift=\xshift] (1.355,4.855) rectangle (2.895,5.145);
        \draw[black,rounded corners,densely dotted, yshift=\yshift, xshift=\xshift] (1.355,5.255) rectangle (2.895,5.545);
        \draw[black,rounded corners,fill=green, yshift=\yshift, xshift=\xshift] (1.35,4.85) rectangle (2.9,5.15) node[pos=.5] {\footnotesize 100\%};
        \draw[black,rounded corners,fill=green, yshift=\yshift, xshift=\xshift] (1.35,5.25) rectangle (2.9,5.55) node[pos=.5] {\footnotesize 100\%};
        \draw[black,rounded corners,fill=white, yshift=\yshift, xshift=\xshift] (3.5,4.75) rectangle (5.25,6.1);
        \node[text width=1cm, text height=1cm, yshift=\yshift, xshift=\xshift] at (4.53,6.25) {\footnotesize $t=5$};
        \draw[black,rounded corners,densely dotted, yshift=\yshift, xshift=\xshift] (3.605,5.255) rectangle (5.15,5.545);
        \draw[black,rounded corners,densely dotted, yshift=\yshift, xshift=\xshift] (3.605,4.855) rectangle (5.15,5.145);
        \draw[black,rounded corners,fill=orange, yshift=\yshift, xshift=\xshift] (3.6,5.25) rectangle (4.55,5.55) node[pos=.5] {\footnotesize 50\%};
        \draw[black,rounded corners,fill=orange, yshift=\yshift, xshift=\xshift] (3.6,4.85) rectangle (4.55,5.15) node[pos=.5] {\footnotesize 50\%};
        \draw[black,rounded corners,fill=white, yshift=\yshift, xshift=\xshift] (5.75,4.75) rectangle (7.5,6.1);
        \node[text width=1.5cm, text height=1cm, yshift=\yshift, xshift=\xshift] at (7.0,6.25) {\footnotesize $t=12$};
        \draw[black,rounded corners,densely dotted, yshift=\yshift, xshift=\xshift] (5.855,5.255) rectangle (7.4,5.545);
        \draw[black,rounded corners,densely dotted, yshift=\yshift, xshift=\xshift] (5.855,4.855) rectangle (7.4,5.145);
        \draw[black,rounded corners,fill=orange, yshift=\yshift, xshift=\xshift] (5.85,5.25) rectangle (6.8,5.55) node[pos=.5] {\footnotesize 50\%};
        \draw[black,rounded corners,fill=orange, yshift=\yshift, xshift=\xshift] (5.85,4.85) rectangle (6.8,5.15) node[pos=.5] {\footnotesize 50\%};
        \draw [black] plot [smooth, tension=1, yshift=\yshift, xshift=2px] coordinates { (2.125,4.65) (2.1,4.2) (1.8,4.0) (1.8,3.5)};
        \draw [black] plot [smooth, tension=1, yshift=\yshift, xshift=\xshift] coordinates { (4.375,4.65) (4.4,4.325) (3.8,3.8) (3.9,3.45)};
        \draw [black] plot [smooth, tension=1, yshift=\yshift, xshift=\xshift] coordinates { (6.625,4.65) (6.6,4.2) (6.15,3.9) (6.1,3.5)};
    \end{tikzpicture}
    \caption{Snapshots of reachable sets for a USV model over the course of a waypoint-tracking mission.
    Possible trajectories (orange) are sampled from the initial state set (black) and bounded by the RSOAs (blue).
    The RSOAs never intersect with the unsafe region (red), so the USV is guaranteed to be safe, but a 50\% decrease in thruster effectiveness causes the USV to slow down.
    The RSOAs accurately capture the behavior of the system despite the thruster malfunction.
    }
    \label{fig:results:reachable_sets_sim_sym}
% \end{figure}
% \begin{figure}[t]
    \centering
    \includegraphics[width=\columnwidth,clip,trim={0 0 30 10}]{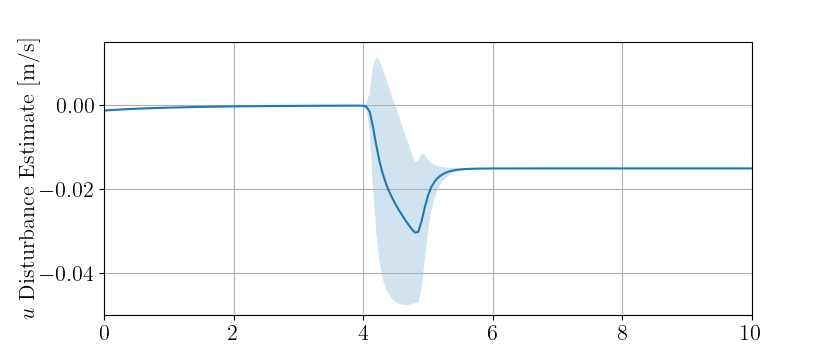}
    \caption{Estimates $(\hat{\bm{\mu}}_t)_{1}$ (line) and $(\hat{\bm{\sigma}}_t)_{1}$ (shaded) for the experiment shown in \cref{fig:results:reachable_sets_sim_sym}.
    The MHE estimates the bias and covariance to characterize the disturbance (thruster malfunction) affecting the USV.}
    \label{fig:results:disturbance_plot}
\end{figure}

\let\thefootnote\relax\footnote{\scriptsize Distribution Statement A. Approved for public release: distribution unlimited. }
\cref{fig:results:reachable_sets_sim_sym} shows an experiment where the USV is commanded to follow the trackline between two waypoints, shown as a dotted black line.
At $t=0$, the USV is behaving as expected, travelling at $\SI[per-mode=symbol]{0.5}{\meter\per\second}$ to the right along the trackline.
Simulated trajectories (orange) are propagated forward from the initial state set (black), and are contained within the RSOAs (blue) over a time horizon of $\SI{2.5}{\second}$.
The RSOAs do not intersect with the unsafe regions (red) $\SI{0.5}{\meter}$ away from the trackline, so the USV is guaranteed to be safe over the time horizon.
At $t=4$, a disturbance is introduced, causing both the USV's left and right thrusters to operate at 50\% effectiveness.
\cref{fig:results:disturbance_plot} shows that until $t=4$, the estimated disturbance in the $u$ dynamics was near $0$, but the MHE registers the loss of control effectiveness as a disturbance, which is captured by the estimates $(\hat{\bm{\mu}}_t)_{1}$ (line) and $(\hat{\bm{\sigma}}_t)_{1}$ (shaded region).
Notice that as the system experiences a new disturbance, the RSOAs inflate in response to the increased uncertainty captured by $\hat{\bm{\sigma}}_t$.
This phenomenon is reflected in the RSOAs calculated at $t=5$, which are stretched horizontally.
At $t=12$, the disturbance estimate has stabilized and the reachable sets accurately capture the sampled trajectories under the new operating conditions.

\subsubsection{Asymmetric Thruster Failure}
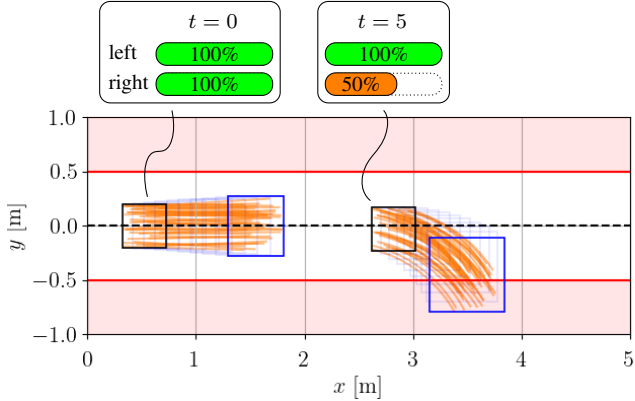
\begin{figure}[t]
    \centering
    \begin{tikzpicture}
        \usetikzlibrary{decorations.pathreplacing}
        \pgfmathsetmacro{\yshift}{-22px}
        \pgfmathsetmacro{\xshift}{18px}
        \node[anchor=south west,inner sep=0] at (0,0) {\includegraphics[width=\columnwidth, clip, trim={5 0 30 20}]{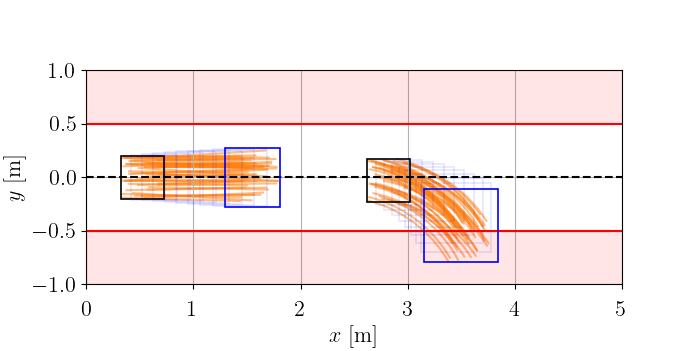}};
        % \draw [thick, decorate,decoration={brace,amplitude=10pt}, yshift=\yshift, xshift=\xshift] (0.6,6.2) -- (7.5,6.2) node [black,midway,yshift=16pt] {\small Thruster Health};
        \draw[black,rounded corners,fill=white, yshift=\yshift, xshift=\xshift] (0.6,4.75) rectangle (3.0,6.1);
        \node[text width=1cm, text height=1cm, yshift=\yshift, xshift=\xshift] at (2.28,6.25) {\footnotesize $t=0$};
        \node[text width=1cm, text height=1cm, yshift=\yshift, xshift=\xshift] at (1.22,5.83) {\footnotesize left};
        \node[text width=1cm, text height=1cm, yshift=\yshift, xshift=\xshift] at (1.22,5.4) {\footnotesize right};
        \draw[black,rounded corners,densely dotted, yshift=\yshift, xshift=\xshift] (1.355,4.855) rectangle (2.895,5.145);
        \draw[black,rounded corners,densely dotted, yshift=\yshift, xshift=\xshift] (1.355,5.255) rectangle (2.895,5.545);
        \draw[black,rounded corners,fill=green, yshift=\yshift, xshift=\xshift] (1.35,4.85) rectangle (2.9,5.15) node[pos=.5] {\footnotesize 100\%};
        \draw[black,rounded corners,fill=green, yshift=\yshift, xshift=\xshift] (1.35,5.25) rectangle (2.9,5.55) node[pos=.5] {\footnotesize 100\%};
        \draw[black,rounded corners,fill=white, yshift=\yshift, xshift=\xshift] (3.5,4.75) rectangle (5.25,6.1);
        \node[text width=1cm, text height=1cm, yshift=\yshift, xshift=\xshift] at (4.53,6.25) {\footnotesize $t=5$};
        \draw[black,rounded corners,densely dotted, yshift=\yshift, xshift=\xshift] (3.605,5.255) rectangle (5.145,5.545);
        \draw[black,rounded corners,densely dotted, yshift=\yshift, xshift=\xshift] (3.605,4.855) rectangle (5.15,5.145);
        \draw[black,rounded corners,fill=green, yshift=\yshift, xshift=\xshift] (3.6,5.25) rectangle (5.15,5.55) node[pos=.5] {\footnotesize 100\%};
        \draw[black,rounded corners,fill=orange, yshift=\yshift, xshift=\xshift] (3.6,4.85) rectangle (4.55,5.15) node[pos=.5] {\footnotesize 50\%};
        % \draw[black,rounded corners,fill=white, yshift=\yshift, xshift=\xshift] (5.75,4.75) rectangle (7.5,6.1);
        % \node[text width=1.5cm, text height=1cm, yshift=\yshift, xshift=\xshift] at (6.8,6.25) {\footnotesize $t=12.5$};
        % \draw[black,rounded corners,densely dotted, yshift=\yshift, xshift=\xshift] (5.855,5.255) rectangle (7.4,5.545);
        % \draw[black,rounded corners,densely dotted, yshift=\yshift, xshift=\xshift] (5.855,4.855) rectangle (7.4,5.145);
        % \draw[black,rounded corners,fill=orange, yshift=\yshift, xshift=\xshift] (5.85,5.25) rectangle (6.8,5.55) node[pos=.5] {\footnotesize 50\%};
        % \draw[black,rounded corners,fill=orange, yshift=\yshift, xshift=\xshift] (5.85,4.85) rectangle (6.8,5.15) node[pos=.5] {\footnotesize 50\%};
        \draw [black] plot [smooth, tension=1, yshift=\yshift, xshift=2px] coordinates { (2.125,4.65) (2.1,4.2) (1.8,4.0) (1.8,3.5)};
        \draw [black] plot [smooth, tension=1, yshift=\yshift, xshift=\xshift] coordinates { (4.375,4.65) (4.4,4.325) (4.1,3.8) (4.2,3.45)};
    \end{tikzpicture}
    \caption{Snapshots of reachable sets over the course of a waypoint-tracking mission where the USV's right thruster malfunctions.
    The disturbance from the thruster malfunction is incorporated into the RSOAs, allowing the analysis to predict a possible collision with the unsafe region.}
    \label{fig:results:reachable_sets_sim_asym}
\end{figure}

\cref{fig:results:reachable_sets_sim_asym} shows a scenario similar to the one discussed in \cref{sec:results:sim:sym}: the USV is commanded to follow the dotted trackline at $\SI[per-mode=symbol]{0.5}{\meter\per\second}$.
In this experiment however, only the USV's right thruster experiences a malfunction.
The asymmetry in the system's control effectiveness is too much for the controller to handle, causing the vehicle to steer to the right at $t=5$.
The RSOAs accurately reflect the bias in the yaw rate, correctly predicting that the USV is going to veer off course.
Because the RSOAs calculated at $t=5$ intersect with the unsafe region, safety cannot be guaranteed over the $\SI{2.5}{\second}$ time horizon.

% \cref{fig:results:disturbance_plot} shows the MHE estimates of $\bm{\mu}_t$ (line) and $\bm{\sigma}_t$ (shaded) for the yaw rate over the same experiment.
% Critically, the MHE recognizes that the right thruster is decreasing in power throughout the experiment as reflected in the estimate of $\bm{\mu}_t$.
% Similarly, the increased uncertainty in the system's dynamics caused by the sudden failure causes the estimate of $\bm{\sigma}_t$ to increase, as is reflected by the wider bounds of the shaded region.
% As shown in \cref{fig:results:reachable_sets_sim_asym}, the end result is that the MHE enables an accurate prediction of the future reachable sets.

% \begin{figure}[t]
%     \centering
%     \includegraphics[width=\columnwidth]{figures/results/disturbance_est_r2.png}
%     \caption{Estimates of $\bm{\mu}_t$ (line) and $\bm{\sigma}_t$ (shaded).
%     The MHE estimates the bias and covariance to characterize the difference between the nominal system and the actual system reflected in the data.}
%     \label{fig:results:disturbance_plot}
% \end{figure}

\begin{figure}[t]
    \centering
    \includegraphics[width=\columnwidth]{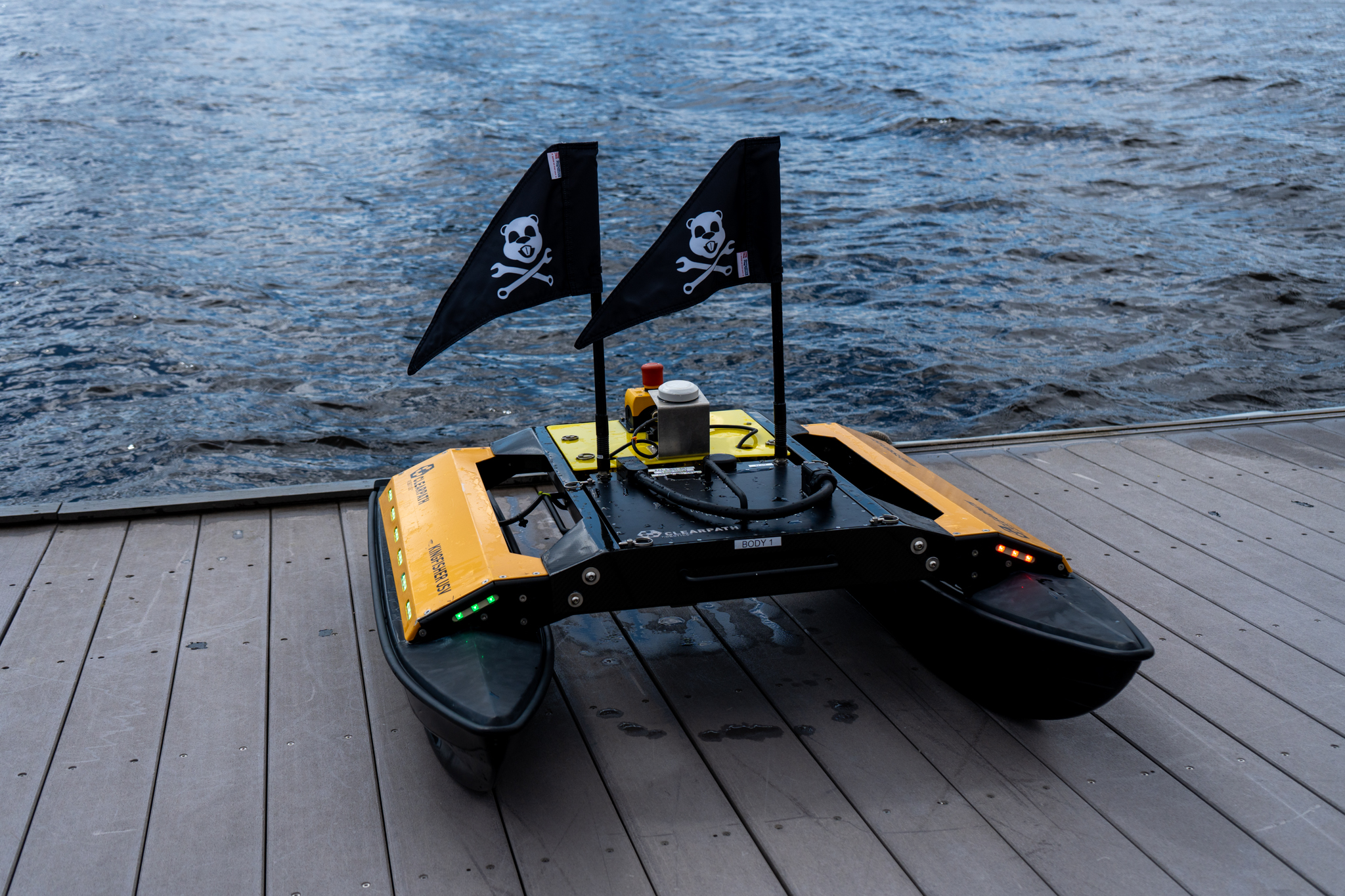}
    \caption{Clearpath Robotics Heron® USV at the MIT Sailing Pavilion.}
    \label{fig:results:heron}
    \vspace{-12pt}
\end{figure}

\begin{figure}[t]
    \centering
    \begin{tikzpicture}
        \usetikzlibrary{decorations.pathreplacing}
        \pgfmathsetmacro{\yshift}{-45px}
        \pgfmathsetmacro{\xshiftone}{-10px}
        \pgfmathsetmacro{\xshifttwo}{54px}
        \node[anchor=south west,inner sep=0] at (0,0) {\includegraphics[width=\columnwidth, clip, trim={5 60 30 80}]{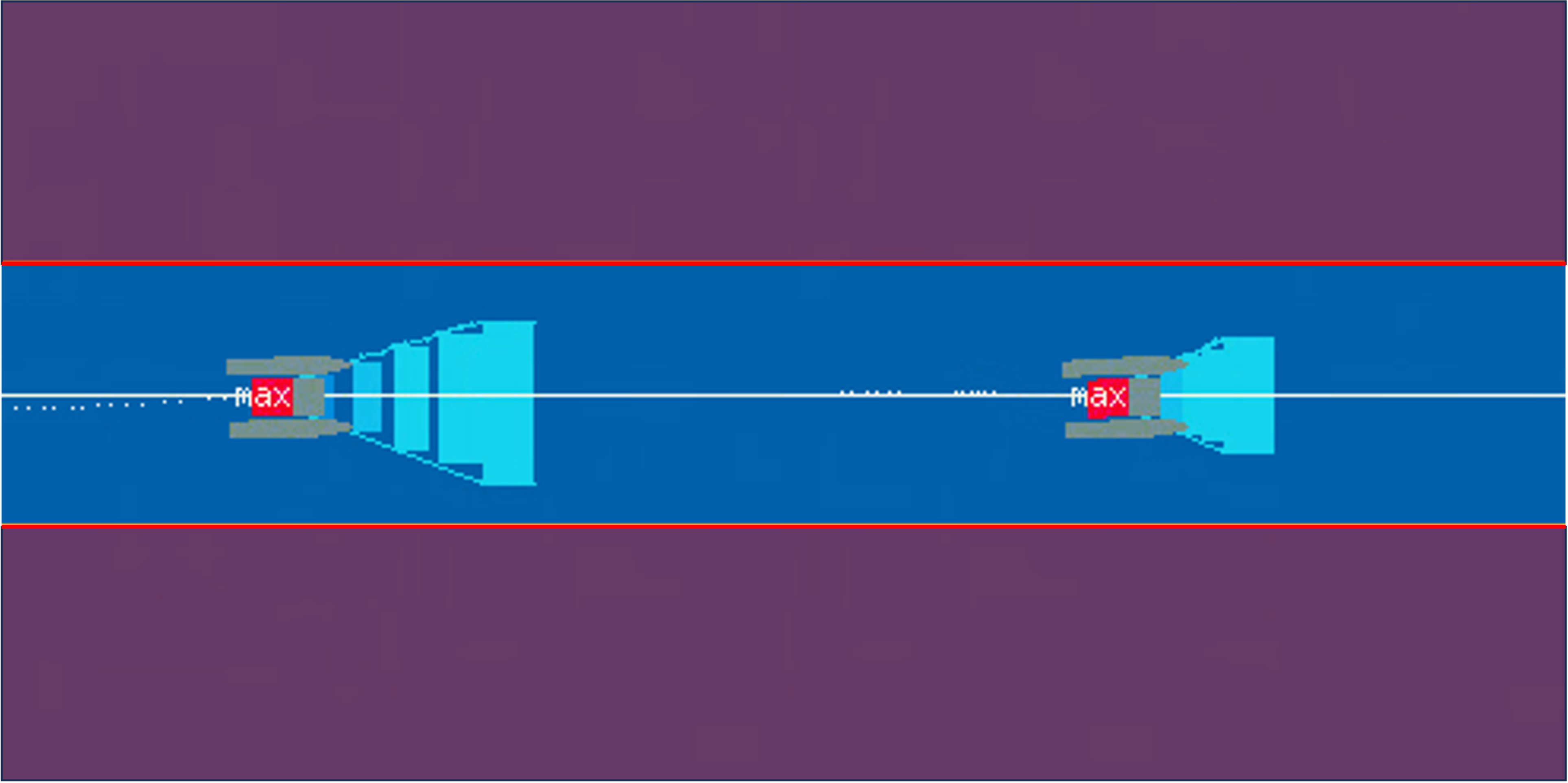}};
        % \draw [thick, decorate,decoration={brace,amplitude=10pt}, yshift=\yshift, xshift=\xshift] (0.6,6.2) -- (7.5,6.2) node [black,midway,yshift=16pt] {\small Thruster Health};
        \draw[black,rounded corners,fill=white, yshift=\yshift, xshift=\xshiftone] (0.6,4.75) rectangle (3.0,6.1);
        \node[text width=1cm, text height=1cm, yshift=\yshift, xshift=\xshiftone] at (2.28,6.25) {\footnotesize $t=0$};
        \node[text width=1cm, text height=1cm, yshift=\yshift, xshift=\xshiftone] at (1.22,5.83) {\footnotesize left};
        \node[text width=1cm, text height=1cm, yshift=\yshift, xshift=\xshiftone] at (1.22,5.4) {\footnotesize right};
        \draw[black,rounded corners,densely dotted, yshift=\yshift, xshift=\xshiftone] (1.355,4.855) rectangle (2.895,5.145);
        \draw[black,rounded corners,densely dotted, yshift=\yshift, xshift=\xshiftone] (1.355,5.255) rectangle (2.895,5.545);
        \draw[black,rounded corners,fill=green, yshift=\yshift, xshift=\xshiftone] (1.35,4.85) rectangle (2.9,5.15) node[pos=.5] {\footnotesize 100\%};
        \draw[black,rounded corners,fill=green, yshift=\yshift, xshift=\xshiftone] (1.35,5.25) rectangle (2.9,5.55) node[pos=.5] {\footnotesize 100\%};
        \draw[black,rounded corners,fill=white, yshift=\yshift, xshift=\xshifttwo] (3.5,4.75) rectangle (5.25,6.1);
        \node[text width=1cm, text height=1cm, yshift=\yshift, xshift=\xshifttwo] at (4.53,6.25) {\footnotesize $t=40$};
        \draw[black,rounded corners,densely dotted, yshift=\yshift, xshift=\xshifttwo] (3.605,5.255) rectangle (5.145,5.545);
        \draw[black,rounded corners,densely dotted, yshift=\yshift, xshift=\xshifttwo] (3.605,4.855) rectangle (5.15,5.145);
        \draw[black,rounded corners,fill=orange, yshift=\yshift, xshift=\xshifttwo] (3.6,5.25) rectangle (4.25,5.55) node[pos=.5] {\footnotesize 30\%};
        \draw[black,rounded corners,fill=orange, yshift=\yshift, xshift=\xshifttwo] (3.6,4.85) rectangle (4.25,5.15) node[pos=.5] {\footnotesize 30\%};
        % \draw[black,rounded corners,fill=white, yshift=\yshift, xshift=\xshift] (5.75,4.75) rectangle (7.5,6.1);
        % \node[text width=1.5cm, text height=1cm, yshift=\yshift, xshift=\xshift] at (6.8,6.25) {\footnotesize $t=12.5$};
        % \draw[black,rounded corners,densely dotted, yshift=\yshift, xshift=\xshift] (5.855,5.255) rectangle (7.4,5.545);
        % \draw[black,rounded corners,densely dotted, yshift=\yshift, xshift=\xshift] (5.855,4.855) rectangle (7.4,5.145);
        % \draw[black,rounded corners,fill=orange, yshift=\yshift, xshift=\xshift] (5.85,5.25) rectangle (6.8,5.55) node[pos=.5] {\footnotesize 50\%};
        % \draw[black,rounded corners,fill=orange, yshift=\yshift, xshift=\xshift] (5.85,4.85) rectangle (6.8,5.15) node[pos=.5] {\footnotesize 50\%};
        \draw [black, line width=0.25mm] plot [smooth, tension=1, yshift=\yshift, xshift=\xshiftone] coordinates { (2.125,4.65) (2.1,4.2) (1.8,4.0) (1.8,3.5)};
        \draw [black, line width=0.25mm] plot [smooth, tension=1, yshift=\yshift, xshift=\xshifttwo] coordinates { (4.375,4.65) (4.4,4.325) (4.1,3.8) (4.2,3.45)};
        \draw [white, |-|, line width=0.25mm] (0.3,0.3) -- (1.83,0.3);
        \node[white] at (1.065,0.5) {$\SI{10}{\meter}$};
    \end{tikzpicture}
    \caption{Hardware experiment with symmetric thruster malfunctions.
    RSOAs (cyan) are smaller after the disturbance is introduced, correctly predicting the lower speed of the USV.
    Because the RSOAs do not intersect with the unsafe region (purple), safety is guaranteed.}
    \label{fig:results:reachable_sets_hardware_sym}
\end{figure}
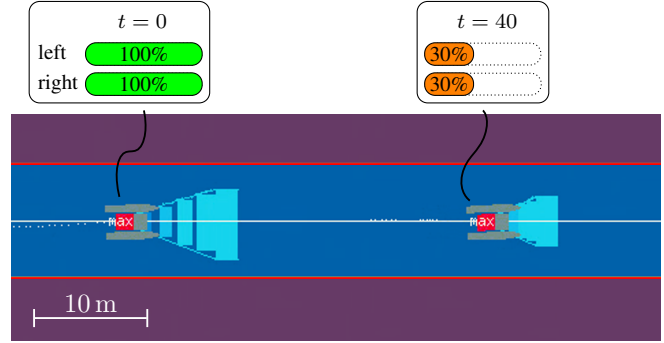

% \begin{figure}[t]
%     \centering
%     % \captionsetup[figure]{aboveskip=0pt,belowskip=-4pt}
%     \begin{subfigure}[t]{0.48\columnwidth}
%         \includegraphics[width=\columnwidth]{figures/results/left.png}
%         \caption{Fully-functioning vehicle with 20 reachable sets extending over a time horizon of 8 s.}
%         \label{fig:results:hardware:nominal}
%     \end{subfigure}~
%     \begin{subfigure}[t]{0.48\columnwidth}
%         \includegraphics[width=\columnwidth]{figures/results/right.png}
%         \caption{Vehicle with both thrusters reduced to 10\% effectiveness.
%         Reachable sets accurately predict slower forward motion.}
%         \label{fig:results:hardware:busted}
%     \end{subfigure}~
%     \caption{Reachable sets predict decrease in forward speed following symmetric thruster failure on a Heron USV.}
%     \label{fig:results:hardware}
%     % \vspace{-16pt}
% \end{figure}

\subsection{Hardware Experiments} \label{sec:results:hardware}
To test the ability of our approach to handle real-world disturbances, we deployed \cref{alg:safety_cert} using a Clearpath Robotics® Heron USV shown in \cref{fig:results:heron}.
The Heron USV is a  1.35$m$ $\times$ 0.98$m$ $\times$ 0.32$m$ catamaran-style vehicle with parallel differential thrusters\cite{heron2020}.
The Heron USV's sensor package includes a compass magnetometer, IMU, GPS module and antenna, WiFi antenna, and RF antenna. 
The Heron USV has two onboard computers for managing high-frequency operations (e.g., motor control) and lower frequency operations (e.g., waypoint following).
These computers communicate with a command station equipped with an Intel® Core™ i7 CPU running at 2.7 GHz.
While the command station is used to run the online computation of our approach in the following experiment, future work will move the calculation onboard the vessel.
For the hardware experiment, we used system identification to estimate a linear set of matrices that represent \cref{eqn:heron_hydrodynamics} and designed a robust servomechanism LQR (RSLQR) as specified in~\cite{Lavretsky.Wise2012_1}.

\let\thefootnote\relax\footnote{\scriptsize Distribution Statement A. Approved for public release: distribution unlimited.}
\cref{fig:results:reachable_sets_hardware_sym} shows the results of the hardware experiment.
Similarly to the experiment described in \cref{sec:results:sim:sym}, the Heron USV is commanded to follow a trackline between two waypoints.
% and train a neural network (NN) control law using imitation learning with state-action pairs generated by a PID controller.
Initially, the USV behaves as expected: about 40\% of its maximum power is applied to each thruster to travel at $\SI[per-mode=symbol]{1}{\meter\per\second}$.
As the USV performs the mission, we induce a malfunction in both thrusters, reducing their effectiveness to 30\% of the commanded thruster output.
The controller attempts to overcome the reduced thruster effectiveness by increasing the commanded power to 100\% ; however, given the degraded thruster health, the Heron USV can only  maintain a forward velocity of $\SI[per-mode=symbol]{0.7}{\meter\per\second}$.
The effect of this disturbance appears in the RSOAs shortly after it is applied: the RSOAs are compressed toward the vehicle, thus reflecting the slower forward velocity.
At each instance in \cref{fig:results:reachable_sets_hardware_sym}, we calculate 20 RSOAs (cyan) over a time horizon of $\tau_r = \SI{8}{\second}$ with an average total computation time of 1.03~$\pm$~$\SI{0.67}{\milli\second}$.

\section{Conclusion}
This paper introduces an online safety certification method that leverages moving horizon estimation (MHE) and forward reachability analysis to predict the future states of a system despite unknown disturbances.
Reachability analysis is often computationally expensive and assumes knowledge of the system's true dynamics; these trade-offs prohibit such methods from being used on real robotics systems.
This method employs $\tt{jax\_verify}$ to construct linear bounds on a computational graph representation of the system's estimated dynamics, which includes the nominal system dynamics and a bias estimate from MHE.

Future work includes extending this method with online system learning techniques to improve the bias estimation throughout the reachability horizon, allowing us to handle more complex, time-varying disturbances in a less conservative way.
Additionally, exploring more rigorous methods of including multi-loop control architectures, e.g., waypoint following, would enable more complex reachable set behaviors and improve this method's safety verification accuracy.
\let\thefootnote\relax\footnote{\scriptsize Distribution Statement A. Approved for public release: distribution unlimited. }
%\balance
\bibliographystyle{IEEEtran}
\bibliography{refs}
\end{document}